\documentclass[preprint,12pt]{elsarticle}

\usepackage{amssymb}

\usepackage{amsmath,amssymb,graphicx,cases}
\usepackage{epsfig}
\usepackage{textcomp}
\usepackage{epstopdf}
\usepackage{float}
\usepackage{braket}
\usepackage[normalem]{ulem}
\usepackage[utf8]{inputenc}
\biboptions{sort&compress}

\usepackage[normalem]{ulem}
\usepackage{dsfont}
\newcommand{\stkout}[1]{\ifmmode\text{\sout{\ensuremath{#1}}}\else\sout{#1}\fi}

\usepackage{fixmath}

\usepackage{color}
\definecolor{Blue}{rgb}{0.00, 0.00, 1.00}
\definecolor{Red}{rgb}{1.00, 0.00, 0.00}
\definecolor{Green}{rgb}{0.00, 0.60, 0.00}
\definecolor{Magenta}{rgb}{0.90, 0.00, 0.90}

\newcommand{\nn}{\nonumber}
\newcommand{\be}{\begin{equation}}
\newcommand{\ee}{\end{equation}}
\newcommand{\bea}{\begin{eqnarray}}
\newcommand{\eea}{\end{eqnarray}}

\usepackage{multirow}
\usepackage{float}
\usepackage{caption, subcaption}

\usepackage[colorlinks=true, urlcolor=blue, anchorcolor=blue, citecolor=blue,filecolor=blue,linkcolor=blue,menucolor=blue]{hyperref}

\journal{Physica A}

\begin{document}

\begin{frontmatter}



\title{Macroscopic fluctuation theory of local time in lattice gases}


%

\author[label1]{Naftali R. Smith}
\ead{naftalismith@gmail.com}
\affiliation[label1]{organization={Department of Environmental Physics, Blaustein Institutes for Desert Research, Ben-Gurion University of the Negev},
            addressline={Sede Boqer Campus},
            postcode={8499000},
            country={Israel}}

\author[label2]{Baruch Meerson}
\ead{meerson@mail.huji.ac.il}
\affiliation[label2]{organization={Racah Institute of Physics, Hebrew University of Jerusalem},
            addressline={Jerusalem},
            postcode={91904},
            country={Israel}}

\begin{abstract}

The local time in an ensemble of particles measures the amount of time the particles
spend in the vicinity of a given point in space. Here we study fluctuations of 
the empirical time average $R= T^{-1}\int_{0}^{T}\rho\left(x=0,t\right)\,dt$ of the density $\rho\left(x=0,t\right)$ at the origin (so that $R$ is the local time spent at the origin, rescaled by $T$) for an initially uniform one-dimensional diffusive lattice gas. We consider both the quenched and annealed initial conditions  and employ the Macroscopic Fluctuation Theory (MFT). For a gas of non-interacting random walkers (RWs) the MFT yields exact
large-deviation functions of $R$, which are closely related to the ones recently obtained by Burenev \textit{et al.} (2023) using microscopic calculations  for non-interacting Brownian particles. Our MFT calculations, however, additionally yield the most likely history of the gas density $\rho(x,t)$ conditioned on a given value of $R$.
Furthermore, we calculate the variance of the local time fluctuations for arbitrary particle- or energy-conserving diffusive lattice gases. Better known examples of such systems include the simple symmetric exclusion process, the Kipnis-Marchioro-Presutti model and the symmetric zero-range process. Our results for the non-interacting RWs can be readily extended to a step-like initial condition for the density.

\end{abstract}

%
%
%
%
%

\end{frontmatter}



\section{Introduction}


Fluctuations in spatially extended macroscopic systems are a central paradigm of interest in statistical mechanics. Macroscopic fluctuations can be quantified by considering different observables, such as the current of mass or energy, activity, \textit{etc}. One useful measure of fluctuations is the ``local time'', \textit{i.e.} the total time spent by the systems's particles at  or around a specified point in space.
The local time has many applications in different fields. For example,
consider chemical or biological reactions, in which a receptor's activity is proportional to the time during which the reactants stay in its vicinity. Then the yield of the product is given by the local time spent by the reactants at the location of the receptor \cite{Wilemski1973,Benichou2005,Doi1975,Temkin1984}.
The local time can also be viewed as the occupation (or residence) time within a given spatial domain, in the limit where the size of the domain vanishes \cite{Levy1940, Redner2001,Pal2019, AKM19, TouchetteMinimalModel, Touchetteoneparticle, SM23}.

The local time has been studied extensively for single-particle systems
\cite{MajumdarSNC2002,SabhapanditS2006,Kishore2020,Carmi2010,Louchard1984, Csorgo1999,Grebenkov2007, Comtet2002,PalA2019,SK21}. Very recently Burenev \textit{et al.} \cite{Burenev} considered a many-body system consisting of noninteracting Brownian particles on the line, whose initial density is uniform on the negative half-line  and zero on the positive one. They studied the probability distribution of the local time at the origin, and focused on the large-deviation regime and on the  (ever-lasting) effect of the type of initial condition -- quenched 
 (that is deterministic) or annealed (random).

In this work, we 
 extend the results of Ref.~\cite{Burenev}  in two directions:

(i) By applying the macroscopic fluctuation theory (MFT) (see Ref. \cite{MFTreview} for a review),
  to the gas of noninteracting random walkers (RWs) on a lattice, we calculate exact
large-deviation functions of $R$ and show that they are closely related to the ones calculated by Burenev \textit{et al.} (2023) using a microscopic approach. In addition we study, in the quenched and annealed cases, the optimal (\textit{i.e.} most likely) history of the gas density, which dominates the probability of observing a given value of the local time $R$. The optimal histories, or paths, are interesting quantities because they provide important insights into the physical mechanisms behind large deviations. The optimal paths can be observed in simulations
\cite{Hurtado2009,Hurtado2011,Lecomte2012,Bunin2012,HMS2019,HMS2021,HM2023} and, in principle, in experiments. In particular, we find that, in the annealed case, the optimal gas density history obeys a time-reversal symmetry  around $t=T/2$.

(ii) Employing the MFT, we calculate the variance of $R$ for a broad class of \emph{interacting} particle- or energy-conserving diffusive lattice gases with arbitrary diffusivity and mobility \cite{Spohn,Liggett,KL,Krapivskybook}. Among the better known examples of such gases are the simple symmetric exclusion process (SSEP), the Kipnis-Marchioro-Presutti (KMP) model \cite{Spohn,Liggett,KL,KMP} and the symmetric zero-range process (ZRP) \cite{Spohn,Liggett,KL,Krapivskybook,Evans,void}. Additional examples include  the Katz-Lebowitz-Spohn (KLS) model \cite{KLS,Baek2017}, the Kob-Andersen kinetically-constrained model \cite{KobAndersen1993,Arita2018}, the repulsion process \cite{K2013}, a family of ``strong particle" models, the facilitated exclusion process, the exclusion process with avalanches, and other models, see Ref.~\cite{Gabrieli2018} and references therein.

The MFT is based on a saddle-point evaluation of the path integral which provides a coarse-grained description of the (noisy) dynamics of the gas density. The ensuing minimization procedure gives rise to saddle-point equations whose solution yields the optimal gas density history.
We assume an initially uniform density on the entire real line. This setting is slightly simpler than the step-function initial density profile considered in Ref.~\cite{Burenev}. We show however that, for the non-interacting RWs, the two local-time distributions for the two initial conditions are simply related.

The rest of the paper is organized as follows. Section \ref{sec:model} starts with a well-known coarse-grained description of diffusive lattice gases \cite{Spohn} and presents the MFT formulation of the local time statistics problem for such gases.
In Sec.~\ref{sec:RWs} we determine the full probability distribution of the local time for noninteracting random walkers (RWs), as well as the optimal histories as described above, for the quenched and annealed initial conditions, and compare our results  for the distribution of local time with those of Ref.~\cite{Burenev}. In Sec.~\ref{sec:interacting} we calculate the variance of the distribution of local time  for interacting lattice gases with arbitrary diffusivity and mobility.
In Sec.~\ref{sec:discussion} we briefly summarize and discuss our results.

\medskip

\section{Model, basic definitions and governing equations}
\label{sec:model}

A rigorous coarse-grained description of stochastic dynamics of diffusive lattice gases is provided by the fluctuating hydrodynamics formalism.  It involves a Langevin equation which effectively accounts, on large scales and at long times, for the contribution of the shot noise to the particle flux  \cite{Spohn}. In one dimension this equation reads
\begin{equation}
    \label{continuity}
    \partial_t \rho + \partial_x j = 0,
\end{equation}
where
\begin{equation}
    \label{langevin}
    j = -D(\rho)\partial_x \rho - \sqrt{\sigma(\rho)} \, \eta.
\end{equation}
Here $\rho(x,t)$ and $j(x,t)$ are the density of particles and current respectively, and $\eta=\eta(x,t)$ is a white, Gaussian noise, satisfying
\begin{equation}
    \label{noise_term}
    \left<\eta(x,t)\right> = 0, \quad \left<\eta(x,t)\eta(x',t')\right> = \delta(x-x')\delta(t-t') .
\end{equation}
The transport coefficients $D(\rho)$ and $\sigma(\rho)$ are the diffusivity and mobility of the gas, respectively, which are determined by the microscopic dynamics of the specific model. Table \ref{table:Dsigma} presents these transport coefficient, as well as the free energy density $F(\rho)$, which is discussed below, for the RWs,  the SSEP, the KMP and the ZRP.

\begin{table}[t]
\label{Table_models}
\renewcommand{\arraystretch}{1.2}
\begin{tabular}{|c|c|c|c|}
\hline
 ~Model ~ &  ~$D(\rho)$ ~ & $\sigma(\rho)$ & ~$F(\rho)$  \\
\hline
RWs      &  1    &      $2\rho$             &  $\rho\ln \rho - \rho$\\
\hline
SSEP     &  1     &  ~$2\rho(1-\rho)$ ~    &   ~$\rho\ln \rho  +(1-\rho)\ln(1-\rho)$ ~\\
\hline
KMP     &  1     &   ~$2\rho^2$ ~     &   ~$-\ln \rho $ ~\\
\hline
ZRP     &  $\alpha'(\rho)$     &   ~$2 \alpha(\rho)$ ~     &   ~$\int^{\rho} \ln \alpha(\rho)\,d\rho$ ~\\
\hline
\end{tabular}
\caption{Transport coefficients $D(\rho)$ and $\sigma(\rho)$ and  the free energy density $F(\rho)$ for the non-interacting RWs and for three interacting particle systems: the SSEP, the KMP model and the ZRP. Here $\alpha(\rho)$
is the microscopic rate of the ZRP, and it is assumed that $\alpha'(\rho)>0$.}
\label{table:Dsigma}
\end{table}


We suppose that at $t=0$ the gas occupies the entire $x$-axis, and the gas density $\rho(x,t=0)=n_0$ is uniform. We will consider both quenched, and annealed initial conditions.
Following Ref. \cite{Burenev}, we are interested in the full probability distribution $\mathcal{P}\left(R\right)$  of the ``local time" (rescaled by $T$), that is the empirical time average of the density at a point which we can choose to be the origin:
\begin{equation}\label{averagedef}
R = \frac{1}{T} \int_0^T \rho(x=0,t)\,dt\,.
\end{equation}
At long times $T \gg 1$ the noise term in Eq.~\eqref{langevin} becomes effectively weak, as shown in  \ref{app:MFT}.  Therefore, $\mathcal{P}\left(R\right)$ can be approximately determined via a saddle-point evaluation of the path integral corresponding to Eqs.~\eqref{continuity}, \eqref{langevin} and (\ref{averagedef}). The action minimization procedure, which we perform explicitly in  \ref{app:MFT}, yields the MFT  equations which can be presented in a Hamiltonian form. In rescaled variables $t/T \to t$, $x/\sqrt{T} \to x$, they read
\begin{eqnarray}
  \partial_t q &=& \partial_{x}\left[D\left(q\right)\partial_{x}q-\sigma\left(q\right)\partial_{x}p\right]\,,\label{qeqGen}\\[1mm]
\partial_{t}p&=&-D\left(q\right)\partial_{x}^{2}p
-\frac{1}{2}\sigma'\left(q\right)\left(\partial_{x}p\right)^{2}
-\Lambda\delta\left(x\right)\,,\label{peqGen}
\end{eqnarray}
where $q(x,t)$ -- a deterministic field -- is the optimal history of the density $\rho(x,t)$, $p(x,t)$ is the ``conjugate momentum" density [which is related to the optimal history of the noise $\xi(x,t)$], and $\Lambda$ is a Lagrange multiplier which accounts for the integral constraint~(\ref{averagedef}) and is ultimately expressed via $R$. As one can see, Eq.~(\ref{peqGen}) contains
a source term $-\Lambda\delta\left(x\right)$ which is specific to the problem of local time statistics.
 A similar delta function source term in the equation for the conjugate momentum field has been encountered before in the context of large deviations of the time-averaged local height of stochastic interfaces \cite{SMV2019}. This term appears there in the framework of the optimal fluctuation method, which is just another name for the MFT that we use in the present work.

To complete the definition of the problem, the MFT equations must be supplemented with boundary conditions in space and time. The spatial boundary conditions are
\be
q\left(\left|x\right|\to\infty,t\right)=n_{0},\quad p\left(\left|x\right|\to\infty,t\right)=0.
\ee
Since the gas density is unspecified at $t=1$, this yields the ``free" boundary condition
\begin{equation}\label{BCp1}
p(x,t=1)=0\,.
\end{equation}
The initial condition at $t=0$ depends on the way in which the initial density is chosen. In the quenched case, the initial density is deterministically uniform, and thus
\begin{equation}\label{quenched0}
q(x,t=0) = n_0\,.
\end{equation}
In the annealed case, the initial condition is random with \emph{average} density $n_0$, corresponding to the assumption that the system had a sufficient time to equilibrate prior to $t=0$.
This leads to the initial condition
\be \label{annealed0general}
p\left(x,t=0\right)=\mathcal{F}'\left[q\left(x,0\right)\right]
\ee
for the MFT equations,  where
\begin{equation}\label{mathcalF}
\mathcal{F}\left(r\right)=F\left(r\right)-F\left(n_{0}\right)-F'\left(n_{0}\right)\left(r-n_{0}\right)\,,
\end{equation}
see \textit{e.g.} Ref.  \cite{DG2009b}. The equilibrium free energy density of the gas $F(r)$  is related to the diffusivity and mobility coefficients via $F''\left(r\right)=2D\left(r\right)/\sigma\left(r\right)$.

Once the MFT problem is solved, the long-time probability distribution $\mathcal{P}\left(R\right)$ can be obtained, in the leading order, by evaluating the action along the optimal path. This leads to a universal large-$T$ scaling behavior
\be
\label{PRScaling}
\mathcal{P}\left(R\right)\sim e^{-\sqrt{T}\,s\left(R\right)}\,.
\ee
The large-deviation function $s(R)$ depends on the type of initial condition -- quenched or annealed -- and, through $D(\rho)$ and $\sigma(\rho)$, on the particular lattice gas model.

For the quenched case, $s(R)$ can be calculated by evaluating the dynamical action $s(R) = s_{\text{dyn}}(R)$ where
\be
\label{sofdxp}
s_{\text{dyn}}(R)=\frac{1}{2}\int_{0}^{1}dt\int_{-\infty}^{\infty}dx \, \sigma\left(q\right)\left(\partial_{x}p\right)^{2}\,.
\ee
In the annealed case, one should also take into account the ``cost" $s_{\text{in}}(R)$ of creating the optimal initial density profile:
\be
s_{\text{in}}(R)=\int_{-\infty}^{\infty}dx\mathcal{F}\left[q\left(x,0\right)\right] \, ,
\label{cost}
\ee
so that the total action is given by
$s = s_{\text{in}} + s_{\text{dyn}}$. Full details of the derivation of the MFT formulation are given in  \ref{app:MFT}.

 In practice, the evaluation of the double integral in Eq.~\eqref{sofdxp} can be difficult. It is often convenient, therefore, to employ a useful shortcut in the form of the relation
\begin{equation}\label{shortcut}
\frac{ds}{dR}=\Lambda\,,
\end{equation}
which follows from the fact that $R$ and $\Lambda$ are conjugate variables,  
see \textit{e.g.} Ref. \cite{Cunden2016}.
\medskip

\section{Complete statistics of the local time for noninteracting RWs}
\label{sec:RWs}

\subsection{General}

For noninteracting RWs, the MFT equations \eqref{qeqGen} and \eqref{peqGen} read
\begin{eqnarray}
  \partial_t q &=& \partial_x^2 q-\partial_x(2 q\partial_x p)\,,\label{qeq}\\[1mm]
  \partial_t p &=& -\partial_x^2 p-(\partial_x p)^2 - \Lambda \delta(x)\,. \label{peq}
\end{eqnarray}
Applying the Hopf-Cole canonical transformation 
$Q=q e^{-p}$ and $P=e^p$ to Eqs.~(\ref{qeq}) and (\ref{peq}), we obtain the linear, decoupled equations
\begin{eqnarray}
  \partial_t Q &=& \partial_x^2 Q + \Lambda Q \delta(x)\,,\label{Qeq}\\
  \partial_t P &=& -\partial_x^2 P- \Lambda P \delta(x)\,. \label{Peq}
\end{eqnarray}
The boundary condition (\ref{BCp1}) becomes
\begin{equation}\label{BCP1}
P(x,t=1) = 1\,.
\end{equation}
The quenched initial condition \eqref{quenched0} becomes
\begin{equation}\label{quenched0Q}
Q(x,t=0) P(x,t=0)=  n_0 \, .
\end{equation}
 By virtue of Eq.~(\ref{mathcalF}) and the relation $F'(r) = \ln r$ (see the first line of Table \ref{table:Dsigma}), the annealed initial condition \eqref{annealed0general} reads
\begin{equation}\label{annealed0}
p(x,t=0) = \ln \frac{q(x,t=0)}{n_0} \,.
\end{equation}
In the Hopf-Cole variables Eq.~\eqref{annealed0} becomes very simple:
\begin{equation}\label{annealed0Q}
Q(x,t=0)=n_0 \, .
\end{equation}

We now proceed to solve the MFT problem in the Hopf-Cole variables. We first determine $P(x,t)$ by solving  Eq.~\eqref{Peq} backwards in time, with ``final" condition \eqref{BCP1}.
Let us denote $f(t)\equiv P(0,t)$, and reverse time, $\tilde{t} = 1-t$, so that Eq.~\eqref{Peq} becomes
\be
\partial_{\tilde{t}}P\left(x,\tilde{t}\right)=\partial_{x}^{2}P\left(x,\tilde{t}\right)+\Lambda f\left(\tilde{t}\right)\delta(x)\label{PeqRev}
\ee
and Eq.~\eqref{BCP1} becomes the initial condition $P\left(x,\tilde{t}=0\right)=1$. Equation~\eqref{PeqRev} is an inhomogeneous linear heat equation with a source term
$\Lambda f\left(\tilde{t}\right)\delta(x)$,
and its formal solution is given by
\begin{equation}\label{Pformal}
P\left(x,\tilde{t}\right)=1+\frac{\Lambda}{\sqrt{4\pi}}\int_{0}^{\tilde{t}}d\tau\,\frac{f\left(\tau\right)}{\sqrt{\tilde{t}-\tau}}e^{-x^{2}/4\left(\tilde{t}-\tau\right)}\,.
\end{equation}
Setting $x=0$ in Eq.~\eqref{Pformal} yields an integral equation for the unknown $f(\tilde{t})$,
\begin{equation}\label{Abel}
\frac{\Lambda}{\sqrt{4\pi}}\int_{0}^{\tilde{t}}d\tau\,\frac{f(\tau)}{\sqrt{\tilde{t}-\tau}}=f\left(\tilde{t}\right)-1\,,
\end{equation}
which is known as the Abel's equation of the second kind.
Its solution, already in the original time $t$, is \cite{Polyanin}
\begin{equation}\label{Abelsol}
f\left(1-t\right)\equiv P\left(x=0,t\right)=e^{\frac{\Lambda^2(1-t)}{4}} \text{erfc}\left(-\frac{\Lambda\sqrt{1-t}}{2}\right)\,,
\end{equation}
where
$\text{erfc}\left(z\right)=1-\left(2/\sqrt{\pi}\right)\int_{0}^{z}e^{-t^{2}}dt$
is the complementary error function.
Plugging Eq.~\eqref{Abelsol} into Eq.~\eqref{Pformal}, one obtains an integral expression for $P(x,t)$. We now continue to solve the MFT problem, first for the annealed case, and then for the quenched case which is a little more complicated.

\subsection{Annealed initial condition}

In the  annealed case, the calculation of $Q(x,t)$ is very similar to that of $P(x,t)$. One notices that Eq.~\eqref{Qeq} with the initial condition~\eqref{annealed0Q} is almost identical to Eq.~\eqref{PeqRev} with the initial condition $P(x,\tilde{t}=0) = 1$. Indeed, for $n_0 = 1$ the two problems are identical, but since the equations are linear, the solution will simply be proportional to $n_0$. In particular, we immediately obtain, by analogy with Eq.~\eqref{Abelsol}, that
\begin{equation}\label{Q0}
Q(x=0,t)=n_{0}e^{\Lambda^{2}t/4}\text{erfc}\left(-\frac{\Lambda\sqrt{t}}{2}\right)\,.
\end{equation}
Using this result, one can obtain an integral expression for $Q(x,t)$ at arbitrary $t$ but, as we now show, this is unnecessary for the purpose of calculating the large-deviation function.
Indeed, from Eqs.~\eqref{Abelsol} and \eqref{Q0} one finds that the density at the origin at times $0 \le t \le 1$ is given by
\be
q(x=0,t)  = Q(0,t) P(0,t) =n_{0}e^{\Lambda^{2}/4}\text{erfc}\left(-\frac{\Lambda\sqrt{1-t}}{2}\right)\text{erfc}\left(-\frac{\Lambda\sqrt{t}}{2}\right).
   \label{q0}
\ee
Plugging Eq.~\eqref{q0} into the rescaled version of Eq.~(\ref{averagedef}), \textit{i.e.} $R=\int_{0}^{1}q\left(x=0,t\right)dt$, we obtain
\begin{equation}\label{RvsLambda}
R(\Lambda)=n_{0}e^{\Lambda^{2}/4}\int_{0}^{1}dt\,\text{erfc}\left(-\frac{\Lambda\sqrt{1-t}}{2}\right)\text{erfc}\left(-\frac{\Lambda\sqrt{t}}{2}\right).
\end{equation}
Evaluating the integral (see  \ref{app:RLambdaIntegral}), we obtain
\begin{equation} \label{RvsLambdaSol}
R\left(\Lambda\right)=\frac{2n_{0}}{\Lambda^{2}}\!\left\{ e^{\Lambda^{2}/4}\!\left(\Lambda^{2}-2\right)\!\left[\text{erf}\left(\frac{\Lambda}{2}\right)+1\right]+\frac{2\Lambda}{\sqrt{\pi}}+2\right\} .
\end{equation}
Now we employ Eq.~(\ref{shortcut}) to calculate the large deviation function $s(R)$. Applying the chain rule, we find
\be
\label{dsdLambdaAnnealed}
\frac{1}{n_{0}}\frac{ds}{d\Lambda}=\frac{\Lambda}{n_{0}}\frac{dR}{d\Lambda}=\frac{e^{\Lambda^{2}/4} \! \left(\Lambda^{4}-2\Lambda^{2}+8\right)  \!  \left[\text{erf}\left(\frac{\Lambda}{2}\right)+1\right] \! + \! \frac{2}{\sqrt{\pi}} \!  \left(\Lambda^{3}-4\Lambda-4\sqrt{\pi}\right)}{\Lambda^{2}}  .
\ee
Integrating Eq.~\eqref{dsdLambdaAnnealed} with respect to $\Lambda$ with the additional condition $s|_{\Lambda=0} = 0$, we obtain
\be
\label{svsLambdaSol}
\frac{s}{n_{0}}=\frac{2\left\{ e^{\Lambda^{2}/4}\left(\Lambda^{2}-4\right)\left[\text{erf}\left(\frac{\Lambda}{2}\right)+1\right]+4\right\} }{\Lambda}+\frac{8}{\sqrt{\pi}}\,.
\ee
Equations~\eqref{RvsLambdaSol} and \eqref{svsLambdaSol} give the rate function $s(R)$ for the annealed initial condition in a parametric form\footnote{\label{simpleconn}One can establish a simple connection between the large-deviation function $s(R)$, as described by Eqs.~\eqref{RvsLambdaSol} and \eqref{svsLambdaSol},  and  the large deviation function for the step initial condition, determined in Ref. \cite{Burenev}. This connection is presented in Sec. \ref{comparison}.}.

\begin{figure*}
\includegraphics[width=.485\textwidth]{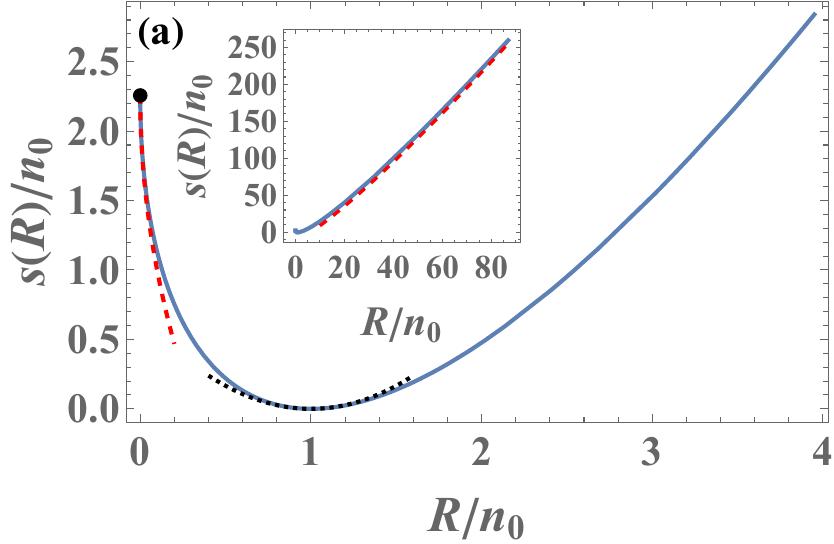}
\hspace{1mm}
\includegraphics[width=.475\textwidth]{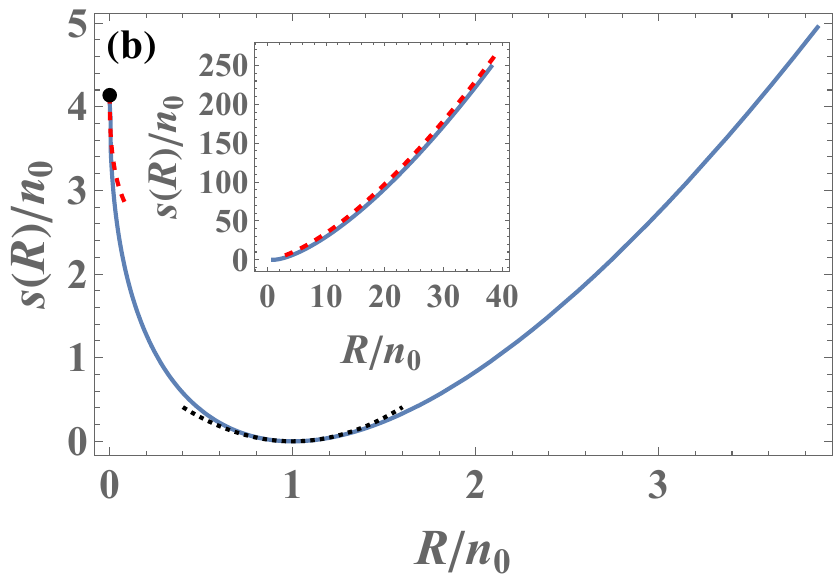}
\caption{Solid lines: $s(R)/n_0$ vs. $R/n_0$ for the annealed (a) and quenched (b) initial conditions. The dashed and dotted lines in the main plots and the dashed lines in the insets correspond to the $R \ll n_0$, $R \simeq n_0$ and $R \gg n_0$ asymptotic behaviors, respectively, see Eqs.~\eqref{sOfRApprox} and \eqref{sOfRApproxQuenched}.}
\label{sOfRannealed}
\end{figure*}
The asymptotic behaviors of $s(R)$ are given by
\be
\label{sOfRApprox}
s\left(R\right)\simeq\begin{cases}
\!4n_0/\sqrt{\pi}- 4\sqrt{R/n_{0}}\,,& R\ll n_{0}\,,\\[2mm]
\!\frac{3\sqrt{\pi}}{8n_{0}}\left(R-n_{0}\right)^{2}\,, & \left|R-n_{0}\right|\!\ll\!n_{0},\\[2mm]
\!R\left[2\sqrt{\ln\left(\frac{R}{4n_{0}}\right)}-\frac{1}{\sqrt{\ln\left(\frac{R}{4n_{0}}\right)}}\right], & R\gg n_{0} \, ,
\end{cases}
\ee
and they are plotted, alongside with the exact $s(R)$, in Fig.~\ref{sOfRannealed} (a).
For completeness, we derive these asymptotics in  \ref{app:Asymptotics}.

 At $R=0$, the first line of Eq.~(\ref{sOfRApprox}) coincides with the action, corresponding to the (two-sided) survival probability of an absorbing wall, located at $x=0$, for the annealed initial condition. The corresponding one-sided survival action, $2n_0/\sqrt{\pi}$, has been known for some time \cite{Blythe2003}.

\begin{figure*}
\includegraphics[width=.48\textwidth]{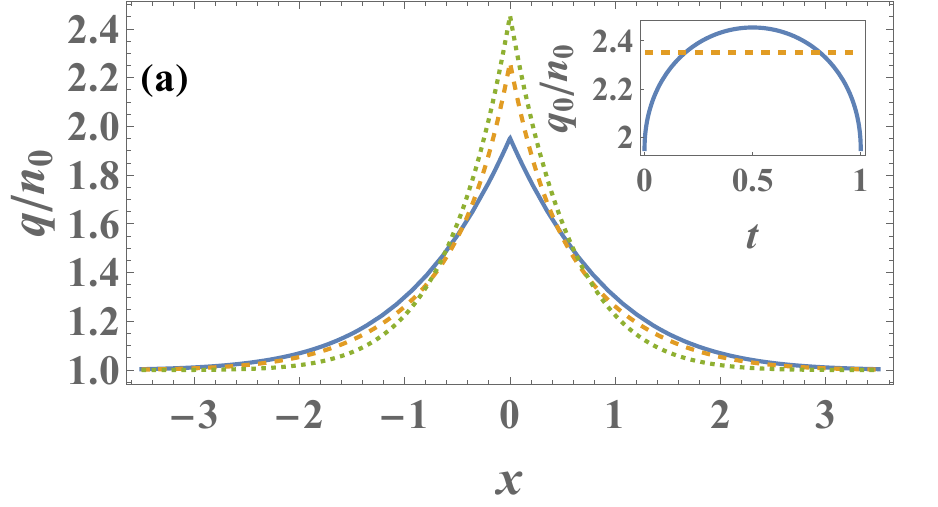}
\hspace{1mm}
\includegraphics[width=.48\textwidth]{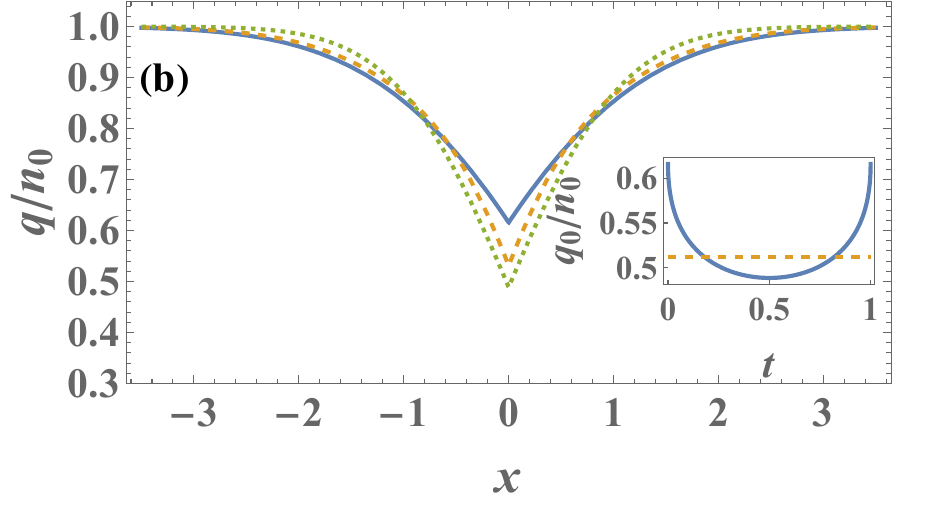}
\caption{The optimal density history of the noninteracting RWs, for the annealed initial condition, as a function of $x$ at times $t=0, 0.1$ and $0.5$ (solid, dashed and dotted lines, respectively) for $\Lambda = 1$ (a) and  $\Lambda = -1$ (b). At times $1/2 \le t \le 1$, $q(x,t)$ is given by the time-reversal symmetry relation $q(x,t) = q(x,1-t)$.  Insets: The optimal densities at the origin, $q_0(t) = q(x=0,t)$ (solid lines) and their time average values $R$ (dashed lines).}
\label{fig:qxt}
\end{figure*}

The optimal history of the gas density, conditioned on a specified $R$, can be determined from the relation $q(x,t) = Q(x,t)P(x,t)$, where $Q$ and $P$ are found as described above. Using the relation $Q(x,t) = n_0 P(x,1-t)$ we find that the optimal history of the density $q(x,t)$ satisfies a time-reversal symmetry $q(x,t) = q(x,1-t)$ around $t=1/2$.
Optimal trajectories for $\Lambda = 1$ and $\Lambda = -1$, corresponding to $R/n_0=2.3520\dots$ and $R/n_0=0.51186\dots$, respectively, are plotted in Fig.~\ref{fig:qxt}.  Notice the nontrivial shape of the optimal  density profile already at $t=0$.  Also noticeable is a corner singularity of the density profile at $x=0$ which is present at all times.

\subsection{Quenched initial condition}

For the quenched case, one solves Eq.~\eqref{Peq} exactly as in the annealed case. However, solving Eq.~\eqref{Qeq} becomes rather more complicated due to the mixed initial condition \eqref{quenched0Q}. This, in turn, makes it difficult to evaluate $R$ as a function of $\Lambda$ directly.

We therefore employ another trick which simplifies the calculation considerably.
We use the property
\bea
\label{dtpq}
\frac{d}{dt}\int_{-\infty}^{\infty}pq\,dx&=&\int_{-\infty}^{\infty}\left(p\partial_{t}q+q\partial_{t}p\right)dx=\nn\\
&=&\int_{-\infty}^{\infty}\left[q\left(\partial_{x}p\right)^{2}-\Lambda\delta\left(x\right)q\right]dx \, ,
\eea
which follows from the MFT equations \eqref{qeq} and \eqref{peq} after performing spatial integrations by parts (using the boundary conditions at $|x| \to \infty$).
From Eq.~\eqref{dtpq}, it follows that the action \eqref{sofdxp} can be recast as
\bea
s&=&\int_{0}^{1}dt\int_{-\infty}^{\infty}dx\,q\left(\partial_{x}p\right)^{2} = \int_{0}^{1}dt\left[\Lambda q\left(0,t\right)+\frac{d}{dt}\int_{-\infty}^{\infty}pq\,dx\right]\nn\\
&=&\Lambda R+\left[\int_{-\infty}^{\infty}pq\,dx\right]_{t=0}^{t=1} = \Lambda R-n_{0}\int_{-\infty}^{\infty}p\left(x,t=0\right)\,dx \, ,
\eea
where in the last equality we used the temporal boundary conditions \eqref{BCp1} and \eqref{quenched0}.
It therefore becomes useful to define the scaled cumulant generating function (SCGF)
\be
\label{mudef}
\mu\left(\Lambda\right)=\Lambda R\left(\Lambda\right)-s\left(\Lambda\right) = n_0 \int_{-\infty}^{\infty}p\left(x,t=0\right)dx\, ,
\ee
which is  the Legendre transform of $s(R)$.

We now proceed to calculate $\mu(\Lambda)$.
For this, it is necessary first to calculate $p(x,t=0) = \ln P(x,t=0)$.
Plugging \eqref{Abelsol} into \eqref{Pformal}, we obtain, at $t=0$,
\be
P\left(x,t=0\right) = 1+\frac{\Lambda}{\sqrt{4\pi}}\int_{0}^{1}d\tau\,\frac{\exp\left(\frac{\Lambda^{2}\tau}{4}-\frac{x^{2}}{4\left(1-\tau\right)}\right)\text{erfc}\left(-\frac{\Lambda\sqrt{\tau}}{2}\right)}{\sqrt{1-\tau}}\,.
\ee
This integral turns out to be solvable, and the result is
\be
\label{PsolInitialQuenched}
P\left(x,t=0\right)=\text{erf}\left(\frac{|x|}{2}\right)+e^{\Lambda\left(\Lambda-2|x|\right)/4} \, \text{erfc}\left(\frac{|x|-\Lambda}{2}\right) \, .
\ee
We were unable to show this fact analytically, but we verified it numerically\footnote{\label{guess}The right-hand side of Eq.~\eqref{PsolInitialQuenched} is closely related to expressions which appear in Ref.~\cite{Burenev}, and this is how we were able to guess the solution to this integral.}. Plugging the logarithm of the expression~(\ref{PsolInitialQuenched}) into Eq.~\eqref{mudef}, we obtain
\bea
\label{muSolQuenched}
\mu\left(\Lambda\right)&=&2n_{0}\int_{0}^{\infty}p\left(x,t=0\right)dx \nn\\
&=&2n_{0}\int_{0}^{\infty}\ln\left[\text{erf}\left(\frac{x}{2}\right)+e^{\Lambda\left(\Lambda-2x\right)/4}\text{erfc}\left(\frac{x-\Lambda}{2}\right)\right]dx.
\eea
Using this result, we obtain $s(R)$ in a parametric form by performing the Legendre transform,
\be
R (\Lambda)=\frac{d\mu}{d\Lambda}= 2n_{0}\int_{0}^{\infty}\frac{e^{-x^{2}/4}\left[2-\sqrt{\pi}e^{(x-\Lambda)^{2}/4}(x-\Lambda)\text{erfc}\left(\frac{x-\Lambda}{2}\right)\right]}{2\sqrt{\pi}\left[\text{erf}\left(\frac{x}{2}\right)+e^{\Lambda(\Lambda-2x)/4}\text{erfc}\left(\frac{x-\Lambda}{2}\right)\right]}dx
\ee
and
\bea
s (\Lambda) &=&\Lambda R-\mu   \nn\\
&=& 2n_{0} \int_{0}^{\infty} \!\! \left\{ \frac{\Lambda e^{-x^{2}/4}\left[2-\sqrt{\pi}e^{(x-\Lambda)^{2}/4}(x-\Lambda)\text{erfc}\left(\frac{x-\Lambda}{2}\right)\right]}{2\sqrt{\pi}\left[\text{erf}\left(\frac{x}{2}\right)+e^{\Lambda(\Lambda-2x)/4}\text{erfc}\left(\frac{x-\Lambda}{2}\right)\right]}\right. \nn\\
&-&  \left.\ln\left[\text{erf}\left(\frac{x}{2}\right)+e^{\Lambda\left(\Lambda-2x\right)/4}\text{erfc}\left(\frac{x-\Lambda}{2}\right)\right]\right\} dx \, .
\eea
An analysis very similar to that performed in \cite{Burenev} yields the asymptotic behaviors
\be
\label{sOfRApproxQuenched}
s\left(R\right)\simeq\begin{cases}
4n_{0}\left(\phi_{\infty}-\sqrt{-\frac{R}{2n_{0}}\ln\frac{R}{n_{0}}}\right), & R\ll n_{0},\\[2mm]
\frac{3\sqrt{\pi}\,n_{0}}{8\left(2-\sqrt{2}\right)}\left(\frac{R}{n_{0}}-1\right)^{2}, & \left|R-n_{0}\right|\ll n_{0},\\[3mm]
\frac{2^{5/2}R^{3/2}}{3^{3/2}n_{0}^{1/2}} \, , & R\gg n_{0},
\end{cases}
\ee
where
$\phi_{\infty}=-\int_{0}^{\infty}\ln\text{erf}\left(z\right)dz=1.03442\dots$. Figure~\ref{sOfRannealed} (b) shows these asymptotics alongside with the exact large-deviation function $s(R)$.

At $R=0$, the first line of Eq.~(\ref{sOfRApproxQuenched}) coincides with the two-sided survival probability of an absorbing wall  at $x=0$ for the quenched initial condition. The twice as small one-sided survival action, $2 n_0 \phi_{\infty}$, was calculated in Ref. \cite{MVK2014}.

\begin{figure*}
\includegraphics[width=.48\textwidth]{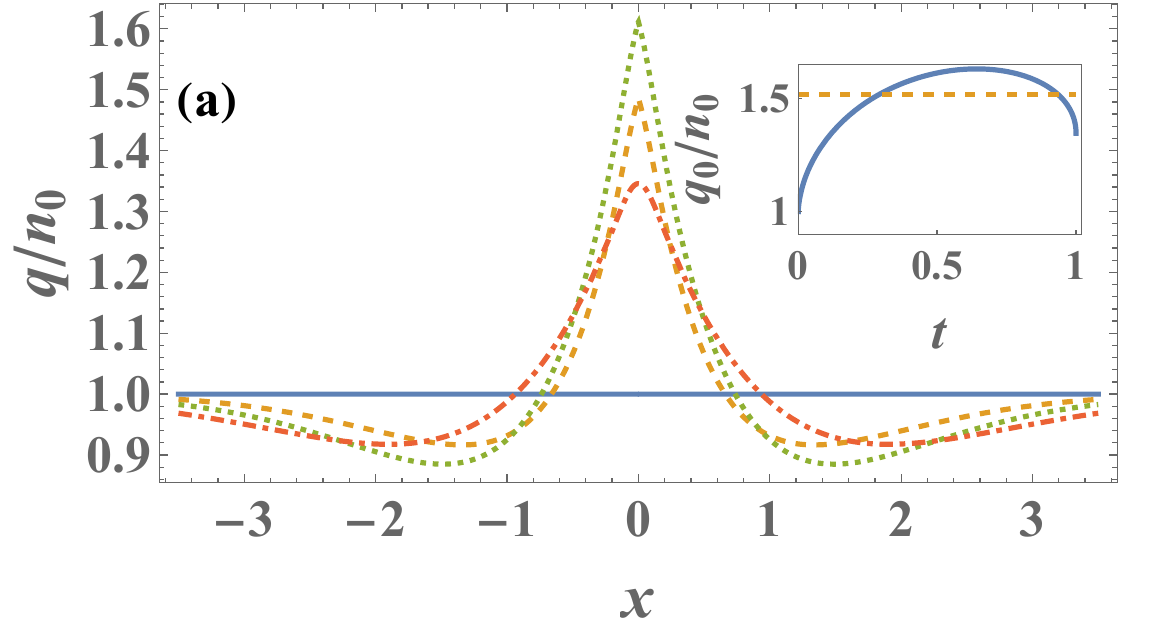}
\hspace{1mm}
\includegraphics[width=.48\textwidth]{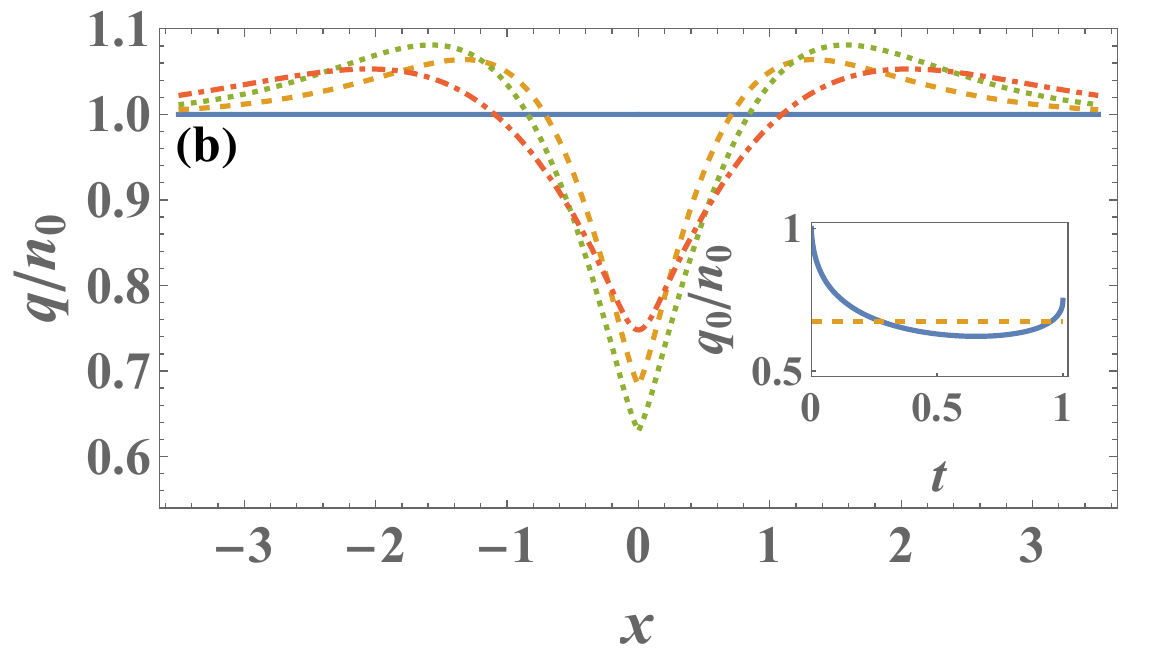}
\caption{ The optimal density history of the noninteracting RWs, for the quenched initial condition, as a function of $x$ at times $t=0, 0.25$, $0.5$ and $1$ (solid, dashed, dotted, and dash-dotted lines, respectively) for $\Lambda = 1$ (a) and  $\Lambda = -1$ (b). Insets: The optimal densities at the origin, $q_0(t) = q(x=0,t)$ (solid lines) and their time average values $R$ (dashed lines).}
\label{fig:qxtquenched}
\end{figure*}

 An analytical calculation of the optimal density history $q(x,t)$ here is more difficult
than in the annealed case. This is because the initial condition \eqref{quenched0Q}
leads to a bulky integral equation which does not appear to be solvable. Fig.~\ref{fig:qxtquenched}
shows two examples of
the optimal density history which we computed numerically for $\Lambda=1$ and $-1$, which correspond to $R/n_0\simeq 1.52$ and
$0.67$, respectively.  We obtained these solutions by numerically
solving Eq.~(\ref{Qeq}) with the initial condition $Q(x,t=0)=n_0/P(x,t=0)$, where $P(x,t=0)$ is given by Eq.~(\ref{PsolInitialQuenched}), and then going back to the original density variable $q(x,t)=Q(x,t) P(x,t)$.

 As one can see, the optimal density history in the quenched case is markedly different from
its counterpart in the annealed case. In particular, the density increase/decrease in the central region around $x=0$ is accompanied by a gas depletion/excess in the surrounding regions. Such a visible redistribution of matter is absent in the annealed setting, see Fig. \ref{fig:qxt}, where the gas has had an infinite time to equilibrate prior to $t=0$. Similar mass redistribution phenomena were observed in the optimal paths of other large-deviation statistics in diffusive lattice gases with quenched initial conditions \cite{void,MS2014}. Finally, in contrast to the annealed case, the optimal density history in the quenched case does not exhibit a time-reversal symmetry. On the other hand, there is still a corner singularity at $x=0$ which persists at all times.

\subsection{Connection with results of Ref.~\cite{Burenev} for step-function initial density}
\label{comparison}

For noninteracting RWs it is straightforward to extend our results for the uniform initial density to a step-function initial density,
\be
\label{stepFunctionRho}
\rho\left(x,t=0\right)=\begin{cases}
n_{1}, & x<0,\\
n_{2}, & x>0,
\end{cases}
\ee
in both the annealed and quenched settings. To this end one can split the local time $R$ into two parts, $R = R_1 + R_2$, where $R_1$ and $R_2$ are the contributions which come from particles whose initial positions were at $x<0$ and $x>0$, respectively.

Since the particles are noninteracting, $R_1$ and $R_2$ are statistically independent.
As a result, the SCGF which describes the distribution of $R$ is given by the sum of the SCGFs of $R_1$ and $R_2$:
\be
\mu(\Lambda) = \mu_1(\Lambda) + \mu_2(\Lambda) \,,
\ee
because the cumulants of $R$ are the sums of the cumulants of $R_1$ and of $R_2$.
In particular, it follows that the SCGF $\mu_{\text{full}}\left(\Lambda\right)$, which describes the homogeneous case $n_1 = n_2$, is related to the SCGF $\mu_{\text{half}}\left(\Lambda\right)$, which describes an initial density vanishing to the right of the origin (\textit{i.e.} $n_2 = 0$), via
\be
\label{fullHalf}
\mu_{\text{full}}\left(\Lambda\right)=2\mu_{\text{half}}\left(\Lambda\right) \,.
\ee
Therefore, using our results above for the homogenous case, one can obtain $\mu_{\text{half}}\left(\Lambda\right)$ and from it the corresponding rate function $s_{\text{half}}\left(R\right)$, and similarly for a general step-function initial condition \eqref{stepFunctionRho}.

The relation \eqref{fullHalf} enables us to directly compare our results for $s(R)$ with those of Ref.~\cite{Burenev} which were obtained for the case $n_2 = 0$.
Indeed, we find that for both the quenched and annealed initial conditions,
\be
\label{muPhi}
\mu\left(\Lambda\right)=-4n_{0}\phi\left(-\frac{\Lambda}{2}\right)\,,
\ee
where the $\mu$'s are the SCGFs $\mu_{\text{full}}$ obtained in the current work, and the $\phi$'s are the functions given in Ref.~\cite{Burenev}.
The minus signs in \eqref{muPhi} are due to different conventions in defining the Legendre transform, and the factors $4$ and $1/2$ are due to the factor of $2$ in Eq.~\eqref{fullHalf} and to our particular choice of diffusion coefficient $D_0 = 1$. With Eq.~(\ref{muPhi}) at hand, one can check
that our exact as well as asymptotic  results for $s(R)$ for the non-interacting RWs agree %
with those of Ref.~\cite{Burenev}.
\medskip

\section{Typical fluctuations of local time in interacting lattice gases}
\label{sec:interacting}

For interacting lattice gases the time-dependent MFT equations can be solved exactly only in exceptional cases \cite{BSM1,Malliketal,BSM2}. Quite often, however, MFT problems are amenable to an asymptotic solution in terms of a systematic expansion of the fields $q(x,t)$ and $p(x,t)$
in the powers of $\Lambda$:
\begin{eqnarray}
  q(x,t) &=& n_0+\Lambda q_1(x,t)+\Lambda^2 q_2(x,t) +\dots \,,\nonumber \\[1mm]
  p(x,t) &=& \Lambda p_1(x,t)+\Lambda^2 p_2(x,t) +\dots\,. \label{perturbseries}
\end{eqnarray}
In practice, this expansion procedure allows one to
obtain first several distribution cumulants,  and it has been used in the studies of
several large-deviation problems for lattice gases, see \textit{e.g.} \cite{KrMe,KMS2014,KMS2015}.

For a uniform initial density, like in the present problem, these calculations become quite simple in the first order of the perturbation theory, where they give the second cumulant of $R$, that is the variance. Indeed, in the first order in $\Lambda$ Eqs.~(\ref{qeq}) and (\ref{peq}) become
\begin{eqnarray}
  \partial_t q_1 &=& D(n_0)\partial_x^2 q_1-\sigma(n_0) \partial^2_x p_1\,,\label{qeq1}\\[1mm]
  \partial_t p_1 &=& -D(n_0)\partial_x^2 p_1- \delta(x)\,. \label{peq1}
\end{eqnarray}
In what follows we will denote for brevity $D(n_0)\equiv D_0$ and $\sigma(n_0)\equiv \sigma_0$. As one can see, Eq.~(\ref{peq1}) is decoupled from Eq.~(\ref{qeq1}), and it can be straightforwardly solved backward in time with the ``initial" condition (\ref{BCp1}). The result is
\be
p_{1}(x,t)=\frac{\Lambda}{2D_{0}}\left[x\,\text{erf}\left(\frac{x}{\sqrt{4D_{0}(1-t)}}\right)+\frac{\sqrt{4D_{0}(1-t)}\,e^{-\frac{x^{2}}{4D_{0}(1-t)}}}{\sqrt{\pi}}-|x|\right]\,.\label{p1sol}
\ee
With this solution at hand, we can determine the action from the linearized version of Eq.~(\ref{sofdxp}),
\be
\label{sofdxplin}
s_{\text{dyn}1}=\frac{1}{2} \Lambda^2 \sigma\left(n_0\right)\int_{0}^{1}dt\int_{-\infty}^{\infty}dx \,\left(\partial_{x}p_1\right)^{2}\,.
\ee
Using Eq.~(\ref{p1sol}), we calculate
\begin{equation}\label{p1xsq}
(\partial_x p_1)^2 = \frac{\Lambda ^2}{{4 D_0^2}} \text{erfc}^2\left(\frac{| x| }{\sqrt{4D_0 (1-t)}}\right)\,,
\end{equation}
substitute  this expression into Eq.~(\ref{sofdxplin}), and evaluate  the double integral. The resulting dynamical
action in terms of $\Lambda$,
\begin{equation}\label{s1vsl}
s_{\text{dyn}1}(\Lambda)= \frac{\left(2-\sqrt{2}\right)\sigma_0\Lambda^2}{3 \sqrt{\pi } D_0^{3/2}}\,,
\end{equation}
combined with the relation $ds/dR = \Lambda$, yields
our final result for the quadratic asymptotic of the action for the quenched initial condition:
\begin{equation}
  s_{\text{quenched}}(R) = \frac{3\sqrt{\pi} D_0^{3/2}}{4 (2-\sqrt{2})\sigma_0} \left(R-n_0\right)^2\,,\label{s1vsRq}
\end{equation}
for $|R-n_0|\ll n_0$. For the noninteracting RWs, with $D_0=1$ and $\sigma_0=2n_0$ (see Table \ref{table:Dsigma}),  the r.h.s. of Eq.~(\ref{s1vsRq}) coincides with the middle line of Eq.~(\ref{sOfRApproxQuenched}), as to be expected.

The (rescaled) variance of $R$ can be immediately read off from Eq.~(\ref{s1vsRq}):
\begin{equation}
  \text{Var}_R^{\text{quenched}} = \frac{2 (2-\sqrt{2}) \sigma_0}{3\sqrt{\pi} D_0^{3/2}}\,.
   \label{varq}
\end{equation}

For the annealed initial condition we should also take into account the free-energy ``cost" of the initial condition, see Eq.~(\ref{cost}). Expanding Eq.~(\ref{mathcalF}) up to, and including, the second order in $\Lambda$, and plugging the result into 
Eq.~(\ref{cost}), we obtain
\begin{equation}\label{sin1}
s_{\text{in}1} (\Lambda) = \frac{1}{2} F''(n_0) \Lambda^2 \int_{-\infty}^{\infty} q_1^2(x,t=0)\,dx\,.
\end{equation}
To determine $q_1(x,t=0)$, we linearize the initial condition \eqref{annealed0general} for the annealed case, which gives
\begin{equation}\label{annealed0lin}
q_1(x,t=0) = \frac{1}{F''(n_0)} p_1(x,t=0)\,.
\end{equation}
Therefore,
\begin{equation}\label{sin1a}
s_{\text{in}1} (\Lambda) =  \frac{\Lambda^2}{2 F''(n_0)} \int_{-\infty}^{\infty} p_1^2(x,t=0)\,dx\,.
\end{equation}
Using the expression (\ref{p1sol}) at $t=0$ and evaluating the integral (\ref{sin1a}), we obtain
\begin{equation}\label{sin2}
 s_{\text{in}1}=\frac{\left(\sqrt{2}-1\right) \sigma_0 \Lambda^2}{3 \sqrt{\pi} D_0^{3/2}}\,,
\end{equation}
 where we used the relation $F''\left(r\right)=2D\left(r\right)/\sigma\left(r\right)$.
The total action in the annealed case is the sum of $s_{\text{dyn}1}$ and $s_{\text{in}1}$, which gives the final result
\begin{equation}\label{s1annvsl}
s_{\text{annealed}1}(\Lambda)=\frac{\sigma_0 \Lambda ^2}{3 \sqrt{\pi } D_0^{3/2}}
\end{equation}
or, in terms of $R$,
\begin{equation}
  s_{\text{annealed}}(R) = \frac{3\sqrt{\pi} D_0^{3/2}}{4 \sigma_0} \left(R-n_0\right)^2\,,\label{s1vsRa}
\end{equation}
for $|R-n_0|\ll n_0$. To remind the reader, $D_0\equiv D(n_0)$ and $\sigma_0\equiv \sigma(n_0)$. For the noninteracting RWs Eq.~(\ref{s1vsRa}) coincides with the middle line of Eq.~(\ref{sOfRApprox}).

The rescaled variance in the annealed case,
\begin{equation}
  \text{Var}_R^{\text{annealed}} = \frac{2 \sigma_0}{3\sqrt{\pi} D_0^{3/2}}\,,
   \label{vara}
\end{equation}
is larger than the variance (\ref{varq}) for the quenched initial condition. This is to be expected: the additional degrees of freedom at $t=0$ facilitate fluctuations.

To conclude this Section, we present our results for the variance in the original, non-rescaled, variables:
\begin{eqnarray}
   \text{Var}_R^{\text{quenched}}  &=& \frac{2 (2-\sqrt{2})  \sigma_0}{3\sqrt{\pi}D_0^{3/2} T^{1/2} }\,, \\
  \text{Var}_R^{\text{annealed}}  &=& \frac{2 \sigma_0}{3\sqrt{\pi} D_0^{3/2}T^{1/2}}\,.
  \label{varphys}
\end{eqnarray}
That is, for all lattice gases, the standard deviation of fluctuations of $R$ around its expected value $\bar{R}=n_0$ goes down with time as $T^{-1/4}$.

\medskip

\section{Summary}
\label{sec:discussion}

To summarize, we studied the full distribution $\mathcal{P}\left(R\right)$ of the local time  (rescaled by $T$) spent at a given spatial point by particles of a one-dimensional lattice gas, assuming an initially uniform gas density. At times $T$ much longer than the characteristic microscopic time of the model, the distribution $\mathcal{P}\left(R\right)$ exhibits a simple and universal scaling behavior \eqref{PRScaling}.  The large deviation function $s(R)$  depends on the underlying microscopic model of the lattice gas only through the transport coefficients $D(\rho)$ and $\sigma(\rho)$. It also depends on the type of initial condition (quenched or annealed) and, in the one-dimensional setting, this dependence persists forever.

For a gas of noninteracting RWs we calculated $s(R)$ exactly  by applying the MFT, thus re-deriving the results of Ref.~\cite{Burenev} who obtained $s(R)$ in a closely-related step-like setting via a microscopic solution of the model of non-interacting Brownian particles. In addition, the MFT equations provide a valuable information about the most likely history of the gas density $q(x,t)$ conditioned on a given value of $R$. This history dominates the contribution to $\mathcal{P}\left(R\right)$. For the RWs we obtained $q(x,t)$ analytically for the annealed initial condition, and numerically for the quenched initial condition. We also showed how our results for $s(R)$ can be straightforwardly extended to the case where the initial densities to the left and right of the origin are not equal.

Further, using the MFT, we studied \emph{typical} fluctuations of $R$ for a general particle- or energy-conserving diffusive lattice gas, for both the quenched and annealed initial conditions.  We did it by calculating the variance of $R$ which depends on the gas properties only through the values of the gas diffusivity and mobility evaluated at the initial (uniform) density.

It would be very interesting, but challenging, to calculate the complete large deviation function $s(R)$ for interacting lattice gases. Recent advances, which exploited the integrability of closely-related MFT equations for the Kipnis-Marchioro-Presutti model  \cite{BSM1,BSM2} and for the SEP \cite{Malliketal}, indicate that this might conceivably be possible in some special cases
 (see also Refs. \cite{KLD21,KLD22,KLD23} in which mathematically similar, or even identical, problems were solved in other physical contexts). If it is not possible to calculate $s(R)$ exactly, one could alternatively probe the $R\to 0$ and $R\to \infty$ tails of $s(R)$ for different lattice gas models, by applying asymptotic methods  to the MFT equations, exploiting the additional small or large parameter $R$.

 It would be also interesting to extend our analysis to other settings and geometries, and also to look for possible singularities in the large-deviation function $s(R)$ in interacting lattice gases.  Such singularities are usually interpreted as dynamical phase transitions. One possible type of singularity arises in lattice gases which exhibit nonlinear diffusion so that the optimal density can vanish identically in some regions of space. Such a singularity was studied in the context of occupation statistics of the ZRP on a ring \cite{AKM19}.

\bigskip

\subsection*{Acknowledgments}
The authors   thank P. L. Krapivsky for a useful discussion and acknowledge support from the Israel Science Foundation (ISF) through Grants No. 2651/23 (NRS) and No. 1499/20 (BM). 

\bigskip

\appendix
\section{Derivation of MFT equations}
\label{app:MFT}

Upon rescaling
$t/T\to t$, $x/\sqrt{T}\to x$, Eqs.~\eqref{continuity} and \eqref{langevin} become
\be
\partial_{t}\rho=\partial_{x}\left[D(\rho)\partial_{x}\rho+T^{-1/4}\sqrt{\sigma(\rho)}\,\eta\right],
\ee
where $\eta$ is also rescaled, by $T^{-3/4}$, and is delta correlated in the rescaled time and space variables.
One therefore finds that, at long times $T\gg1$, the noise term becomes effectively weak.

Let us now derive the MFT equations \eqref{qeq} and \eqref{peq} of the main text.
The probability density of the (dimensionless) Gaussian white noise is
\be
\mathcal{P}\left[\eta\right]\sim\exp\left(-\int_{0}^{1}dt\int_{-\infty}^{\infty}dx\frac{\eta^{2}}{2}\right).
\ee
Expressing $\eta$ through $\rho$ and $j$, we can therefore write the probability of a joint history of the latter two as
$P\left[\rho,j\right]\sim e^{-\sqrt{T}\,s}$
with the action functional
\be
\label{sdef}
s=\int_{0}^{1}dt\int_{-\infty}^{\infty}dx\frac{\left[j+D\left(\rho\right)\partial_{x}\rho\right]^{2}}{2\sigma\left(\rho\right)} \, .
\ee
Exploiting the large parameter $\sqrt{T} \gg 1$, we now evaluate $\mathcal{P}(R)$ by applying the saddle-point approximation to the path integral. Within this framework, $-\ln\mathcal{P}(R) / \sqrt{T}$ is, in the leading order, given by the minimum of the action functional $s$ over all realizations of $\rho$ and $j$, constrained on the initial condition for the density $\rho$, and on a given value of $R$.
We incorporate the latter constraint via a Lagrange multiplier, \textit{i.e.} we minimize the modified action
\be
s_{\Lambda}=s-\Lambda R=\int_{0}^{1}dt\int_{-\infty}^{\infty}dx\left\{ \frac{\left[j+D\left(\rho\right)\partial_{x}\rho\right]^{2}}{2\sigma\left(\rho\right)}
-\Lambda\delta\left(x\right)\rho\left(x,t\right)\right\}\,,
\ee
where $\Lambda$ is a Lagrange multiplier whose value is ultimately set by the constraint $\int_{0}^{1}\rho\left(x=0,t\right)dt=R$.

It is convenient to define a ``potential" $\psi(x,t)$ such that
\be
\rho=\partial_{x}\psi,\quad j=-\partial_{t}\psi \, .
\ee
(The existence of such a potential is guaranteed by the continuity equation.)
Rerwriting the modified action as a functional of $\psi$, we obtain
\bea
s_{\Lambda}&=&\int_{0}^{1}dt\int_{-\infty}^{\infty}dx\left\{ \frac{\left[D\left(\partial_{x}\psi\right)\partial_{x}^{2}\psi-\partial_{t}\psi\right]^{2}}{2\sigma\left(\partial_{x}\psi\right)}-\Lambda\delta\left(x\right)\partial_{x}\psi\left(x,t\right)\right\}  \nn\\
&=&\int_{0}^{1}dt\int_{-\infty}^{\infty}dx\left\{ \frac{1}{2}\sigma\left(\partial_{x}\psi\right)\left(\partial_{x}p\right)^{2}
+\Lambda\delta'\left(x\right)\psi\left(x,t\right)\right\}\,,
\eea
where we integrated by parts and defined the momentum density gradient
\be
\label{pdef}
\partial_{x}p=\frac{D\left(\partial_{x}\psi\right)\partial_{x}^{2}\psi-\partial_{t}\psi}{\sigma\left(\partial_{x}\psi\right)} \, .
\ee
To minimize $s_\Lambda$, we calculate its variation with respect to a small variation of $\psi$,
\bea
&&\delta s_{\Lambda}=\int_{0}^{1}\!dt\int_{-\infty}^{\infty}\!\!\!dx\left\{ -\frac{1}{2}\sigma'\left(\partial_{x}\psi\right)\left(\partial_{x}p\right)^{2}\partial_{x}\delta\psi\right. \nn\\
&&+\left.\left(\partial_{x}p\right)\left[D'\left(\partial_{x}\psi\right)\partial_{x}^{2}\psi\partial_{x}\delta\psi+D\left(\partial_{x}\psi\right)\partial_{x}^{2}\delta\psi-\partial_{t}\delta\psi\right]+\Lambda\delta'\left(x\right)\delta\psi\left(x,t\right)\right\} . \nn\\
\eea
To get rid of the derivatives from the terms proportional to $\delta \psi$, we integrate by parts in time or space. This yields
$\delta s_{\Lambda}=\left.\delta s_{\Lambda}\right|_{\text{bulk}}+\left.\delta s_{\Lambda}\right|_{\text{boundary}}$
where the bulk term is
\bea
\left.\delta s_{\Lambda}\right|_{\text{bulk}}&=&\int_{0}^{1}\!dt\int_{-\infty}^{\infty}\!\!\!dx\left\{ \partial_{x}\left[\frac{1}{2}\sigma'\left(\partial_{x}\psi\right)\left(\partial_{x}p\right)^{2}\right]-\partial_{x}\left[\left(\partial_{x}p\right)D'\left(\partial_{x}\psi\right)\partial_{x}^{2}\psi\right]\right. \nn\\
&+&\left.\partial_{x}^{2}\left[\left(\partial_{x}p\right)D\left(\partial_{x}\psi\right)\right]+\partial_{xt}p+\Lambda\delta'\left(x\right)\right\} \delta\psi\,.
\eea
The boundary terms from the spatial integration by parts all vanish, while the temporal integration by parts yields a term
\be
\label{sBoundary}
\left.\delta s_{\Lambda}\right|_{\text{boundary}}= - \int_{-\infty}^{\infty}dx\left[\partial_{x}p\,\delta\psi\right]_{t=0}^{t=1} \, .
\ee
We now require $\delta s_{\Lambda}$ to vanish for variations $\delta \psi(x,t)$ at arbitrary $-\infty < x < \infty$ and $t>0$.
For the bulk term, this requirement yields the equation
\be
\partial_{t}v=\partial_{x}\left[-D\left(q\right)\partial_{x}v-\frac{1}{2}\sigma'\left(q\right)v^{2}\right]-\Lambda\delta'\left(x\right)
\ee
where $v =\partial_x p$ and $q(x,t)$ is the optimal realization of the density $\rho(x,t)$. After a spatial integration, this equation yields the second MFT equation, Eq.~\eqref{peq} in the main text.
The first MFT equation, Eq.~\eqref{qeq} in the main text, follows from a spatial integration of the definition \eqref{pdef} of the momentum density.

Let us now derive the boundary conditions for the MFT equations. Requiring that the $t=1$ term in Eq.~\eqref{sBoundary} vanish for arbitrary $\delta \psi$, we obtain $\partial_{x}p\left(x,t=1\right)=0$, which after a spatial integration yields Eq.~\eqref{BCp1} of the main text.
The initial condition for the quenched case is particularly simple. Here, $\rho(x,t=0)$ is specified, so for the uniform-density initial condition, considered in the present paper, one simply has $q(x,t=0) = n_0$, which is Eq.~\eqref{quenched0} of the main text.
The annealed case is slightly more tricky. Here, the initial density $\rho_0(x) = \rho(x,t=0)$ is assumed to have equilibrated before time $t=0$, and is therefore randomly sampled from the equilibrium distribution
\be
\mathcal{P}\left[\rho_{0}\left(x\right)\right]\sim e^{-\int_{-\infty}^{\infty}dx\mathcal{F}\left[\rho_{0}\left(x\right)\right]}\,,
\ee
where the function $\mathcal{F}(r)$ is given by Eq.~(\ref{mathcalF}) of the main text. 
As a result, in the minimization problem defined above, one must add to the action $s_\Lambda$ an additional term,
\be
s_{\text{in}}=\int_{-\infty}^{\infty}dx\mathcal{F}\left[\rho\left(x,0\right)\right]
=\int_{-\infty}^{\infty}dx\mathcal{F}\left[\partial_{x}\psi\left(x,0\right)\right] \, ,
\ee
which corresponds to the ``cost'' of the initial condition. The variation of this term is
\be
\delta s_{\text{in}}=\int_{-\infty}^{\infty}dx \, \mathcal{F}'\left[\rho\left(x,0\right)\right]\delta\rho\left(x,0\right)=-\int_{-\infty}^{\infty}dx \, \partial_{x}\left\{ \mathcal{F}'\left[\partial_{x}\psi\left(x,0\right)\right]\right\} \delta\psi\left(x,0\right)\,.
\ee
The initial condition is then obtained by requiring that the sum of $\delta s_{\text{in}}$ and the $t=0$ term in Eq.~\eqref{sBoundary},
\be
\int_{-\infty}^{\infty} \,  dx\partial_{x}\left\{ p\left(x,t=0\right)-\mathcal{F}'\left[\partial_{x}\psi\left(x,0\right)\right]\right\} \delta\psi\left(x,0\right)=0\,,
\ee
vanishes for arbitrary variations $\delta \psi(x,0)$, yielding that
$\mathcal{F}'\left[\partial_{x}\psi\left(x,0\right)\right]-p\left(x,t=0\right)$ is a constant. This constant, however, must vanish, due to the condition $\mathcal{F}'(n_0) = 0$ and the vanishing boundary conditions at $|x|\to\infty$, thus yielding
\be
p\left(x,t=0\right)=\mathcal{F}'\left[\partial_{x}\psi\left(x,0\right)\right]\,,
\ee
which is Eq.~\eqref{annealed0general} of the main text.

Once the MFT problem is solved, the probability of observing a given $R$ is given by $\mathcal{P}\left(R\right)\sim e^{-\sqrt{T}\,s}$, where $s$ is the action \eqref{sdef} evaluated on the solution to the MFT equations which corresponds to the given $R$.
Plugging Eq.~\eqref{pdef} into \eqref{sdef}, one obtains the following expression for $s$:
\be
s=\frac{1}{2}\int_{0}^{1}dt\int_{-\infty}^{\infty}dx \, \sigma\left(q\right)\left(\partial_{x}p\right)^{2}\,,
\ee
which is Eq.~\eqref{sofdxp} in the main text.

\section{Evaluation of the integral \eqref{RvsLambda} which gives $R(\Lambda)$ for the annealed case}
\label{app:RLambdaIntegral}

Let us denote the integral which appears in Eq.~\eqref{RvsLambda} by
\begin{equation}
\label{IAdef}
I\left(A\right)=\int_{0}^{1}dt\,\text{erfc}\left(A\sqrt{1-t}\right)\text{erfc}\left(A\sqrt{t}\right)\,,
\end{equation}
where $A = -\Lambda/2$.
We now take the derivative under the integration sign in \eqref{IAdef} to obtain
\begin{eqnarray}
\label{IprimeA}
I'\left(A\right)&=&\int_{0}^{1}dt\,\left[-\frac{2\sqrt{t}e^{-A^{2}t}\text{erfc}\left(A\sqrt{1-t}\right)}{\sqrt{\pi}}-\frac{2\sqrt{1-t}e^{-A^{2}(1-t)}\text{erfc}\left(A\sqrt{t}\right)}{\sqrt{\pi}}\right]\nonumber\\
&=&\int_{0}^{1}dt\,\left[-2\frac{2\sqrt{t}e^{-A^{2}t}\text{erfc}\left(A\sqrt{1-t}\right)}{\sqrt{\pi}}\right] \nn\\
&=&\frac{2}{A^{3}}\left[\frac{e^{-A^{2}}\left(2A-\sqrt{\pi}\left(A^{2}+1\right)\right)}{\sqrt{\pi}}+\text{erfc}(A)\right] \,,
\end{eqnarray}
where, when moving from the first line to the second line in Eq.~\eqref{IprimeA}, we reversed the time $t \to 1-t$ in the second integral.
Integrating Eq.~\eqref{IprimeA} with respect to $A$ we get
\begin{equation}
I\left(A\right)=2-\frac{2A^{2}\text{erf}(A)-\frac{e^{-A^{2}}\left(\sqrt{\pi}-2A\right)}{\sqrt{\pi}}+\text{erfc}(A)}{A^{2}}
\end{equation}
where we used that $I\left(0\right)=1$ [which follows immediately from \eqref{IAdef}].
And now finally we have [using Eq.~\eqref{RvsLambda}]
\begin{equation}
R\left(\Lambda\right)=n_{0}e^{\frac{\Lambda^{2}}{4}}I\left(-\frac{\Lambda}{2}\right)=\frac{2n_{0}}{\Lambda^{2}}\left\{ e^{\frac{\Lambda^{2}}{4}}\left(\Lambda^{2}-2\right)\left[\text{erf}\left(\frac{\Lambda}{2}\right)+1\right]+\frac{2\Lambda}{\sqrt{\pi}}+2\right\} \,,
\end{equation}
which is Eq.~\eqref{RvsLambdaSol} of the main text.

\section{Asymptotic behaviors of $s(R)$ in the annealed case}
\label{app:Asymptotics}


At $\Lambda \to \infty$, Eqs.~\eqref{RvsLambdaSol} and \eqref{svsLambdaSol} become
$\bar{R} = R / n_0\simeq4/\Lambda^{2}$ and $\bar{s} =s/ n_0\simeq4/\sqrt{\pi}+8/\Lambda$, which together yield the first line in Eq.~\eqref{sOfRApprox} of the main text.
At $\Lambda \to 0$, one finds
$\bar{R}\simeq1+4\Lambda/\left(3\sqrt{\pi}\right)$,
$\bar{s}\simeq2\Lambda^{2}/\left(3\sqrt{\pi}\right)$ which yield the middle line in Eq.~\eqref{sOfRApprox}.
At $\Lambda \to \infty$, Eqs.~\eqref{RvsLambdaSol} and \eqref{svsLambdaSol} become
\bea
\label{RLambdaApprox}
\bar{R} \simeq\left(4-\frac{8}{\Lambda^{2}}\right)e^{\Lambda^{2}/4}, \\
\label{sLambdaApprox}
\bar{s} \simeq\left(4\Lambda-\frac{16}{\Lambda}\right)e^{\Lambda^{2}/4} \, ,
\eea
respectively.
Inverting the relation \eqref{RLambdaApprox}  perturbatively at $R\gg 1$, we find that
$\Lambda\simeq2\sqrt{\ln\left(\frac{\bar{R}}{4}\right)}+\frac{1}{2\ln^{3/2}\left(\frac{\bar{R}}{4}\right)}$.
Using these relations in \eqref{sLambdaApprox}, we find that
\be
\bar{s}\simeq\Lambda\bar{R}-\frac{2\bar{R}}{\Lambda}\simeq2\bar{R}\left[\sqrt{\ln\left(\frac{\bar{R}}{4}\right)}-\frac{1}{2\sqrt{\ln\left(\frac{\bar{R}}{4}\right)}}\right] \, ,
\ee
which is the third line in Eq.~\eqref{sOfRApprox} of the main text. 


%
%





\end{document}